\documentclass[pra,aps,superscriptaddress,twocolumn]{revtex4-1}

\usepackage[colorlinks,bookmarks=false,citecolor=magenta,linkcolor=magenta,urlcolor=
magenta]{hyperref}
\usepackage{amsmath,amssymb}
\usepackage{setspace}%使用间距宏包
\usepackage{graphicx}% Include figure files
\usepackage{float}
\usepackage{dcolumn}% Align table columns on decimal point
\usepackage{bm}% bold math
\usepackage{subfigure}%多图
\usepackage{pifont} %带圈数字
\usepackage{braket}
\usepackage{siunitx}%单位
\usepackage{booktabs}
\usepackage{array}
\usepackage{natbib}
\usepackage{afterpage}
\usepackage{xcolor}

\usepackage[normalem]{ulem} % 提供 \sout 删除线

\definecolor{mypink}{rgb}{1.0, 0.3, 0.6}
\begin{document}

\preprint{APS/123-QED}

\title{Experimental Extraction of Coherent Ergotropy and Its Energetic Cost in a Superconducting Qubit}% Force line breaks with \\

\author{Li Li}
\affiliation{
Beijing National Laboratory for Condensed Matter Physics, 
Institute of Physics, Chinese Academy of Sciences, Beijing 100190, China
}
\affiliation{
School of Physical Sciences, University of Chinese Academy of Sciences, Beijing 100049, China
}

\author{Silu Zhao}
\affiliation{
Beijing National Laboratory for Condensed Matter Physics, 
Institute of Physics, Chinese Academy of Sciences, Beijing 100190, China
}
\affiliation{
School of Physical Sciences, University of Chinese Academy of Sciences, Beijing 100049, China
}
\author{Yun-Hao Shi}
\affiliation{
Beijing National Laboratory for Condensed Matter Physics, 
Institute of Physics, Chinese Academy of Sciences, Beijing 100190, China
}

\author{Kai Xu}

\affiliation{
Beijing National Laboratory for Condensed Matter Physics, 
Institute of Physics, Chinese Academy of Sciences, Beijing 100190, China
}

\affiliation{
Beijing Academy of Quantum Information Sciences, Beijing 100193, China
}
\affiliation{
Hefei National Laboratory, Hefei 230088, China
}

\author{Heng Fan}

\affiliation{
Beijing National Laboratory for Condensed Matter Physics, 
Institute of Physics, Chinese Academy of Sciences, Beijing 100190, China
}
\affiliation{
School of Physical Sciences, University of Chinese Academy of Sciences, Beijing 100049, China
}

\affiliation{
Beijing Academy of Quantum Information Sciences, Beijing 100193, China
}
\affiliation{
Hefei National Laboratory, Hefei 230088, China
}

\author{Dongning Zheng}
\affiliation{
Beijing National Laboratory for Condensed Matter Physics, 
Institute of Physics, Chinese Academy of Sciences, Beijing 100190, China
}

\affiliation{
School of Physical Sciences, University of Chinese Academy of Sciences, Beijing 100049, China
}
\affiliation{
Hefei National Laboratory, Hefei 230088, China
}

\author{Zhongcheng Xiang}
\email{zcxiang@iphy.ac.cn}

\affiliation{
Beijing National Laboratory for Condensed Matter Physics, 
Institute of Physics, Chinese Academy of Sciences, Beijing 100190, China
}
\affiliation{
Hefei National Laboratory, Hefei 230088, China
}

\begin{abstract}

Quantum coherence, encoded in the off-diagonal elements of a system's density matrix, is a key resource in quantum thermodynamics, fundamentally limiting the maximum extractable work known as ergotropy. While previous experiments have isolated coherence-related contributions to work extraction, it remains unclear how coherence can be harnessed in a controllable and energy-efficient manner. Here, we experimentally investigate the role of initial-state coherence in work extraction from a superconducting transmon qubit. By preparing a %range
variety of pure states and implementing three complementary extraction protocols, we reveal how coherence governs the partitioning of ergotropy. We find that the choice of initial state depends on the dominant decoherence channel-energy relaxation or dephasing. By further accounting for thermodynamic costs, we identify optimal initial states that maximize the efficiency. Our results demonstrate that the initial-state design provides a scalable approach to coherence control and advances the development of efficient quantum thermodynamic devices.
\end{abstract}
\maketitle
Quantum batteries (QBs), which are quantum systems designed for energy storage and on-demand release, promise transformative advantages over classical batteries, particularly for scalable, high-efficiency energy transfer~\cite{rmp_campaioli2024,alicki2013,skrzypczyk2014,binder2015,bruschi2015,campaioli2017,ferraro2018,andolina2019,liu2021,moraes2021,dou2022b,dou2022,zhao2022b,mojaveri2023,gyhm2024,lai2024,song2024,elghaayda2025,rossini2025}. A central challenge in QB optimization lies in maximizing the ergotropy, namely the maximum energy a quantum system can deliver to a work reservoir via cyclic unitary operations~\cite{allahverdyan2004}. Recent theoretical advances highlight that quantum coherence~\cite{chitambar2019,streltsov2017} plays a pivotal role in enhancing ergotropy, potentially improving extraction efficiency beyond classical limits~\cite{korzekwa2016,francica2019a,cakmak2020, francica2020,shi2022}. This crucial observation was recently confirmed by a pioneering experiment~\cite{niu2024}, in which the authors used an NV center qubit, gradually dephased a fixed pure state, and observed the corresponding reduction in ergotropy.

However, these findings still leave important questions about how coherence can be effectively harnessed in realistic, controllable settings. In practical systems, decoherence arises from uncontrollable interactions with the environment and represents a fundamentally passive process~\cite{deleon2021, clerk2010}. Although the significance of coherence is well understood, the fact that coherence typically decays passively limits its usefulness for device optimization. In contrast, preparing the initial state of a quantum system is an active process that allows deliberate control and tuning. For quantum energy storage, such active state design is essential to mitigate coherence loss and enhance performance. Consequently, it is necessary to experimentally examine how initial-state coherence impacts work extraction, enabling optimal energy storage strategies. Moreover, the work extraction process consists of a series of unitary operations, each accompanied by thermodynamic cost that can be theoretically quantified~\cite{zheng2016, campbell2017, abah2019, monsel2020}. Incorporating thermodynamic cost into experimental studies and exploring the relationship between coherence and energy efficiency are also key steps toward optimizing the performance of quantum thermodynamics systems. A critical gap remains: How can we actively design initial states with tailored coherence to optimize ergotropy extraction while minimizing thermodynamic costs?

In this work, we bridge this gap by experimentally investigating how initial-state coherencein a superconducting transmon qubit governs ergotropy extraction and thermodynamic efficiency. First, we implement three distinct work extraction protocols, successfully distinguishing between the coherence-independent and coherence-consuming components of ergotropy. Then, by preparing the qubit in various pure states, we explore how the initial state's coherence affects the relative contributions of these two ergotropy components. We further analyze thermodynamic costs and identify optimal states that maximize preparation efficiency and relate extraction efficiency to coherence. Our results highlight coherence as a key factor in quantum energy storage and conversion, and open a new perspective on controlling coherence through initial-state design.
\begin{figure}
    \centering
    \includegraphics[scale = 0.7]{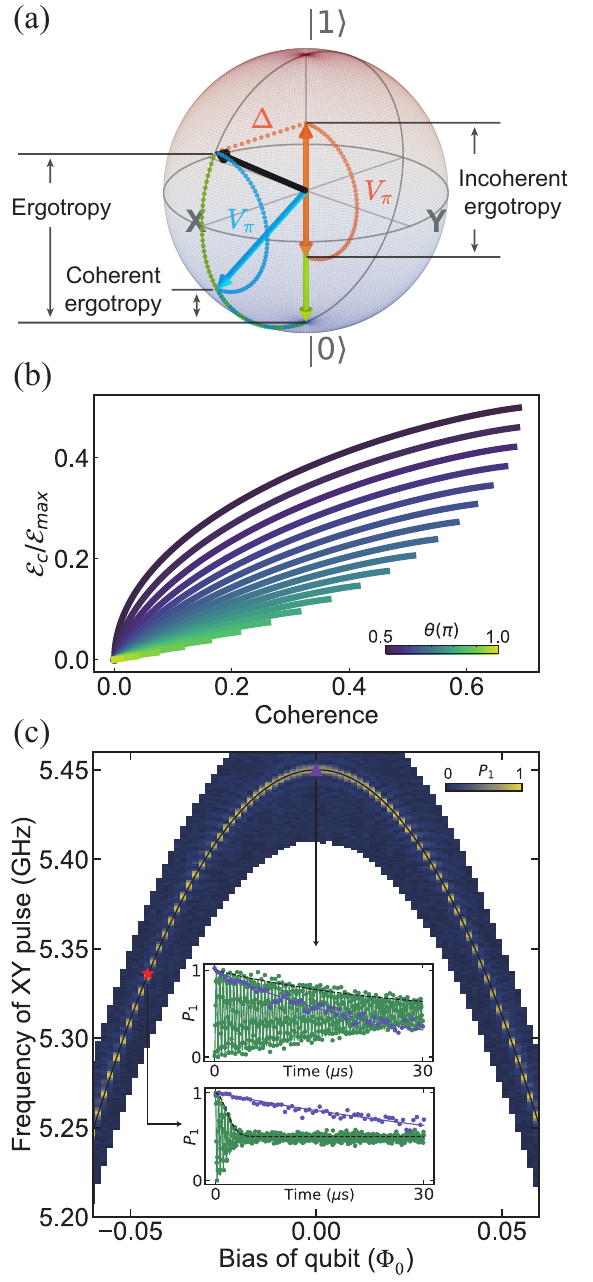}
    \caption{Coherent and incoherent ergotropy in a superconducting qubit.
(a) Bloch sphere representation of ergotropy extraction from a pure state $\rho_S = \ket{\psi(\theta)}\bra{\psi(\theta)}$ (black arrow). The ergotropy $\mathcal{E}(\rho_S)$ corresponds to the green arc to the ground state (passive state). Dephasing $\Delta$ yields the mixed state $\delta_{\rho_S}$ (upward orange arrow), whose ergotropy is the orange arc. The blue arc denotes incoherent ergotropy—the maximal work extractable without consuming coherence—ending in $\sigma_{\rho_S}$. The remainder, coherent ergotropy, is extractable via a unitary $U_c$.
(b) Coherent ergotropy as a function of coherence and Rabi angle $\theta$. It increases with coherence at fixed $\theta$, and depends on $\theta$ due to gate-induced coherence.
(c) Experimental implementation using a transmon qubit. $\Phi_0 = h/2e$ is the flux quantum, where $e$ is the elementary charge and $h$ is Planck’s constant. The flux dependence of the qubit frequency $\omega_q$ shows a sweet spot (purple triangle) with long dephasing time $T_2 = 32.7\mu\mathrm{s}$ and short $T_1 = 25.7\mu\mathrm{s}$, and a detuned point (red star) with $T_2 = 2.2\mu\mathrm{s}$ and $T_1 = 64.5\mu\mathrm{s}$, enabling preparation of $\delta_{\rho_S}$.}
    \label{fig1}
\end{figure}

We begin with the work extraction cycle of a two-level system, governed by the Hamiltonian $H = H_0 + H_d(t)$. The static part is $H_0=-\hbar \omega_q\sigma_z/2$, where $\omega_q$ is the transition frequency between the ground state $\ket{0}$ and the excited state $\ket{1}$ and $\hbar$ is the reduced Planck constant. The time-dependent term $H_d(t)$ represents external driving, implementing unitary operations for state preparation and work extraction. We prepare the initial state as a pure state $\rho_S = \ket{\psi(\theta)}\bra{\psi(\theta)}$, with $\ket{\psi(\theta)} = \cos(\theta/2)\ket{0} + \sin(\theta/2)\ket{1}$, where $\theta$ is the Rabi angle corresponding to a $Y$-axis rotation. This state is illustrated by the black arrow on the Bloch sphere in Fig.~\ref{fig1}(a). In this representation, the energy of state $\rho_S$ (denoted by $E_{\rho_S}$) corresponds to its projection onto the $Z$-axis, while the extractable work is defined as the energy difference between the initial and final states. The ergotropy, the maximum extractable work via unitary evolution $U$, is given by~\cite{allahverdyan2004}
\begin{align}
 \mathcal{E}(\rho_S) = \max_{U} \left( \mathrm{Tr}[\rho_S H_0] - \mathrm{Tr}[U \rho_S U^\dagger H_0] \right).
\end{align}
The optimal $U$ maps $\rho_S$ to its passive state $P_{\rho_S}$, from which no work can be further extracted. For a pure state, this passive state is simply the ground state $\ket{0}$, as shown by the downward green arrow in Fig.~\ref{fig1}(a), so its ergotropy equals its energy. The ergotropy extraction follows the green arc in Fig.~\ref{fig1}(a). 

However, mixed states exhibit fundamentally different behavior. We define the fully dephased state of $\rho_S$ as $\delta_{\rho_S}$, obtained via the dephasing process $\Delta$. This state is shown as the upward orange arrow in Fig.~\ref{fig1}(a), and $\Delta$ is indicated by the orange straight-line trajectory. Its passive counterpart $P_{\delta_{\rho_S}}$ appears as the downward orange arrow. The energy gap between $\delta_{\rho_S}$ and $P_{\delta_{\rho_S}}$—i.e., the ergotropy of $\delta_{\rho_S}$—is strictly smaller than that of $\rho_S$, revealing that coherence, while not affecting energy, determines the amount of extractable work.

The coherence of $\rho_S$ is quantified by the quantum relative entropy~\cite{baumgratz2014}
\begin{align}
C(\rho_S) = D(\rho_S \| \delta_{\rho_S}) = \mathrm{Tr}[\rho_S(\log \rho_S - \log \delta_{\rho_S})],
\end{align}
which represents the ``distance" between $\rho_S$ and its fully dephased counterpart. The component of ergotropy independent of coherence is termed \emph{incoherent ergotropy} $\mathcal{E}_i$. It is equivalently defined as the ergotropy of $\delta_{\rho_S}$ (orange arc in Fig.~\ref{fig1}(a)) or as the maximal work extractable from $\rho_S$ without disturbing its coherence (blue arc). The final state in the latter case, denoted $\sigma_{\rho_S}$, is the lowest-energy state preserving the coherence of $\rho_S$. The operation used to extract the incoherent ergotropy is denoted by $V_\pi$, which corresponds to a $\pi$-rotation around the $X-$axis in the Bloch sphere. The remaining portion of work extractable by consuming coherence is called \emph{coherent ergotropy} $\mathcal{E}_c$, which equals the ergotropy of $\sigma_{\rho_S}$, and the operation is denoted by $U_c$.

For a fixed Rabi angle $\theta$, reducing coherence leads to a monotonic decrease in $\mathcal{E}_c$. A partially dephased state can be written as
\begin{align}
\rho_S^\Delta =
\begin{bmatrix}
\sin^2(\theta/2) & a \\
a^* & \cos^2(\theta/2)
\end{bmatrix},
\end{align}
where $|a| \leq \sin(\theta/2)\cos(\theta/2)$ controls the residual coherence. The coherence $C(\rho_S^\Delta)$ dependence with $a$ quantifies the degree of phase information. Fig.~\ref{fig1}(b) shows that $\mathcal{E}_c$ increases with $C(\rho_S^\Delta)$ for a fixed $\theta$. Besides, Fig.~\ref{fig1}(b) further reveals how $\mathcal{E}_c$ depends on the Rabi angle varying between $\pi/2$ and $\pi$. This effect is nontrivial: the available coherence resource directly stems from the unitary gate used during initialization. In large-scale QBs, such dependencies influence charging power and efficiency. We now investigate this phenomenon using a superconducting transmon qubit~\cite{koch2007}.

To experimentally distinguish coherent and incoherent ergotropy, it is essential to prepare the relevant intermediate quantum states. The most challenging among them is the mixed state $\delta_{\rho_S}$, requiring implementation of $\Delta$: fast suppression of off-diagonal elements while preserving diagonal populations. The mixed-state preparation is closely related to noise engineering and plays an important role in open-system quantum simulation~\cite{harrington2022,mi2024,li2025a,liang2024,tao2024,han2021,barreiro2011,murch2012,brown2022,liu2023a}. In superconducting circuits, one approach introduces auxiliary systems—qubits or resonators—to induce controllable dissipation~\cite{han2021,liu2023a}, but this demands custom hardware. Alternatively, one can apply engineered noise via the qubit’s control lines, e.g., random XY noise~\cite{li2025a} or synthesized low-frequency Z noise~\cite{liang2024,tao2024}. However, XY noise typically drives the system to a unique steady state, unsuitable for our purposes. Z-noise-based dephasing is often limited by device quality, which must ensure spectrally clean relaxation dynamics free of two-level defects~\cite{simmonds2004,lisenfeld2010a,lisenfeld2010,burnett2014,klimov2018,meissner2018}.

In this study, we exploit the qubit’s intrinsic environment to induce dephasing. Figure~\ref{fig1}(c) shows the dependence of a transmon qubit's transition frequency on magnetic flux $\Phi$. The point marked by the purple triangle is the transmon's sweet spot, where $\omega_q^{s}/2\pi = 5.450\,\text{GHz}$ and $\partial \omega_q / \partial \Phi = 0$. This location is immune to flux noise and yields a long dephasing time, $T_2 = 32.7\,\mu\mathrm{s}$. However, it is also closest to the readout resonator frequency, resulting in strong Purcell decay and a short energy relaxation time, $T_1 = 25.7\,\mu\mathrm{s}$. Conversely, the red star marks a flux bias point with the opposite trade-off: $\omega_q/2\pi = 5.336\,\text{GHz}$, $T_1 = 64.5\,\mu\mathrm{s}$ and $T_2 = 2.2\,\mu\mathrm{s}$. These parameters are well suited for implementing $\Delta$ and preparing $\delta_{\rho_S}$. The drive is applied through the XY control line. The drive Hamiltonian takes the form $H_d(t) = g_d v(t)\cos\left(\omega_dt\right) \sigma_y$, where $g_d$ is the drive-qubit coupling strength, $v(t)$ is the time-dependent envelope and $\omega_d$ is the frequency of drive. We take $\omega_d = \omega_q$ in the following experiment.
\begin{figure}[t]
    \centering
    \includegraphics[scale = 0.8]{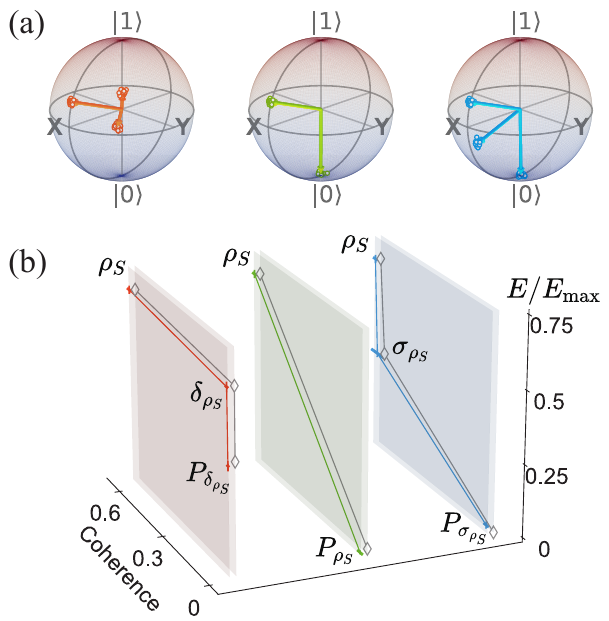}
    \caption{Experimental comparison of three work extraction protocols from a superconducting qubit.
(a) Bloch sphere representation of quantum states under dephasing extraction (orange), direct extraction (green), and sequential extraction (blue). Dots indicate individual state preparations; arrows show average Bloch vectors.
(b) Energy–coherence diagram for the extracted states. Each trajectory illustrates how energy and coherence are reduced under different protocols. The error bars represent the standard error. Gray diamonds represent numeric simulation results accounting for all decoherence effects.} 
    \label{fig2}
\end{figure}
\begin{figure*}[t]
    \centering
    \includegraphics[scale = 0.8]{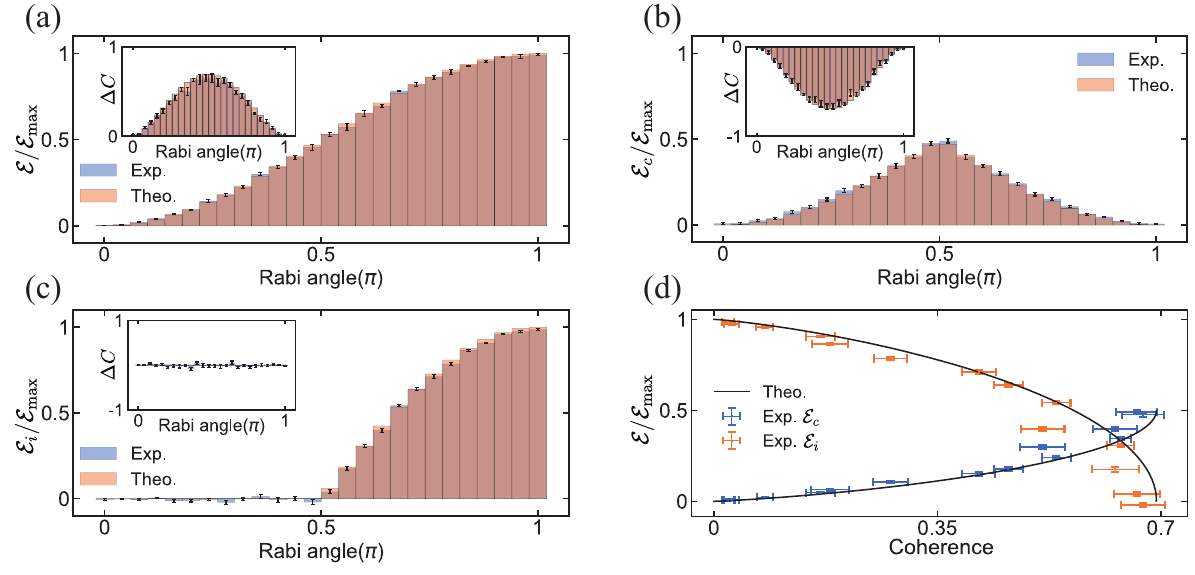}
    \caption{Variation of ergotropy and coherence in the sequential protocol as a function of the Rabi angle $\theta$.
(a) Change in $\mathcal{E}$ and coherence $\Delta C$ of the prepared state $\rho_\theta$ relative to the ground state. The blue bars represent experimental data, and the brown bars represent theoretical calculations.
(b) Coherent ergotropy $\mathcal{E}_c$ versus $\theta$; the inset shows the coherence drop after coherent extraction during $U_c$, indicating full consumption. 
(c) Incoherent ergotropy $\mathcal{E}_i$ as a function of $\theta$, becoming dominant near $\theta = \pi$. The inset shows coherence remaining unchanged during $V_\pi$. 
(d) Initial coherence $C(\rho_\theta)$ plotted against $\mathcal{E}_c$ and $\mathcal{E}_i$, revealing their respective positive and negative correlations with coherence. The black solid line represents the theoretical calculation. All error bars represent the standard error.} 
    \label{fig3}
\end{figure*}

Figure~\ref{fig2} presents the experimental results for three distinct work extraction protocols, which we refer to as dephasing extraction, direct extraction, and sequential extraction. In all cases, the initial state is the pure state $\rho_S = \ket{\psi_0}\bra{\psi_0}$, $\ket{\psi_0} = \left(\sqrt{2}\ket{1} + \ket{0}\right)/\sqrt{3}$, with a Rabi angle $\theta_S = 2\arccos\left({\sqrt{3}/3}\right)$. The final states are reconstructed via quantum state tomography to obtain their full density matrices~\cite{james2001,thew2002}. Figure~\ref{fig2}(a) illustrates the state preparation outcomes on the Bloch sphere. For all protocol, state preparation is repeated twenty times, with the results plotted as individual points. The corresponding average Bloch vectors are shown as arrows. Figure~\ref{fig2}(b) quantifies the relationship between coherence and energy for each state, where all energies are normalized to the maximum energy scale, $E_\mathrm{max} = \hbar \omega_q$.

In the dephasing extraction protocol, the initial state $\rho_S$ is held at the working point for $4\,\mu\text{s}$, during which phase decoherence suppresses all off-diagonal elements, yielding the fully dephased state $\delta_{\rho_S}$. A subsequent application of the $V_{\pi}$ operation transforms $\delta_{\rho_S}$ into its passive state $P_{\delta_{\rho_S}}$. The experimental results for this process are shown in orange in both Figs.~\ref{fig2}(a) (left side) and (b) (left path). The $\mathcal{E}_i$ extracted in this protocol is found to be $0.26\pm0.02$. In the direct extraction protocol, the total ergotropy of $\rho_S$ is extracted via a single unitary operation that is the inverse of the state preparation gate. The corresponding results are shown in green in the center of Figs.~\ref{fig2}(a) and (b). The sequential extraction protocol decomposes the process into two steps. Firstly, a $V_{\pi}$ operation is applied to $\rho_S$, producing the state $\sigma_{\rho_S}$ and extracting the $\mathcal{E}_i$, measured to be $0.33\pm0.02$. Secondly, the remaining coherence is consumed by transforming $\sigma_{\rho_S}$ into its passive state $P_{\sigma_{\rho_S}}$ by $U_c$, extracting the $\mathcal{E}_c$ which is $0.34\pm0.01$. These operations are represented by the blue states and trajectories on the right side of Figs.~\ref{fig2}(a) and (b). The discrepancy in $\mathcal{E}_i$ between the dephasing and sequential protocols arises from unavoidable decay of the population terms during the dephasing process $\Delta$. To model this quantitatively, we simulate the full ergotropy extraction dynamics using the measured decoherence parameters of the superconducting qubit and numerically solve the master equation with qutip~\cite{johansson2012,johansson2013}. The results, plotted as gray diamond markers in Fig.~\ref{fig2}(b), show good agreement with experimental data. These findings demonstrate the experimental distinction between coherent and incoherent ergotropy on a superconducting qubit and confirm the dependence of coherent ergotropy on the amount of quantum coherence.

We prepare initial states with varying Rabi angles $\theta$ for the sequential extraction protocol and denote the extracted changes in energy and coherence during the process. Figure~\ref{fig3}(a) shows the state preparation results as $\theta$ varies from $0$ to $\pi$, the corresponding operation is denoted as $R_Y(\theta)$. The energy change from the ground state to $\rho_\theta$ is $\sin^2\!\left(\theta/2\right)\hbar\omega_q$, and the coherence is $-2\sin^2\!\left(\theta/2\right)\log\left(\sin(\theta/2)\right) - 2\cos^2\!\left(\theta/2\right)\log\left(\cos(\theta/2)\right)$ (see Supplemental Material for details). When $\theta < \pi/2$, the state $\rho_\theta$ itself is passive with the same coherence. Therefore, the $\mathcal{E}_i$ is zero and all extractable work is coherent. For $\pi/2 < \theta < \pi$, we apply a $V_{\pi}$ operation to prepare $\sigma_{\rho_\theta}$ and then perform coherent work extraction to prepare $P_{\sigma_{\rho_\theta}}$. Figure~\ref{fig3}(b) displays the coherent ergotropy versus $\theta$, which reaches a maximum of 0.5 at $\theta = \pi/2$. The inset of Fig.~\ref{fig3}(b) shows the coherence difference before and after coherent extraction, indicating full coherence consumption. Subtracting $\mathcal{E}_c$ from total work yields the $\mathcal{E}_i$ as a function of $\theta$, shown in Fig.~\ref{fig3}(c). At $\theta = \pi$, all work is incoherent. Figure~\ref{fig3}(d) plots the initial coherence against coherent (incoherent) ergotropy, confirming the positive (negative) correlation. This guides optimal charging of QBs: in systems dominated by phase decoherence such as superconducting, NV centers, or spins qubits, energy should be stored incoherently by preparing $\theta = \pi$. When dynamical decoupling~\cite{uhrig2007,cywinski2008,cai2012,xu2012,guo2018a,miao2020,huang2021} extends dephasing time $T_2\simeq 2\,T_1$, coherent storage with $\theta = \pi/2$ is preferable. 
\begin{figure}[t]
    \centering
    \includegraphics[scale = 1.2]{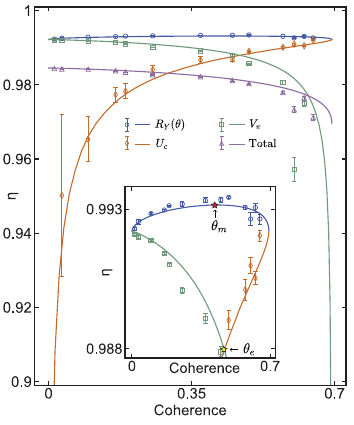}
    \caption{Thermodynamic efficiencies $\eta$ of state preparation $R_Y(\theta)$, coherent extraction $U_c$, incoherent extraction $V_\pi$, and total extraction versus coherence $C(\rho_\theta)$. The inset marks a local maximum preparation efficiency at $\theta_m \approx 0.742\pi$ with $C_m \approx 0.432$ and equal extraction efficiencies at $\theta_e \approx 0.722\pi$ with $C_e \approx 0.469$. The error bars represent the standard error.} 
    \label{fig4}
\end{figure}

Finally, we analyze the thermodynamic cost and efficiency associated with each step of the sequential extraction protocol. For a unitary quantum gate of duration $\tau$, the thermodynamic cost can be expressed as~\cite{abah2019,monsel2020,deffner2021,hu2022,ge2023}
\begin{align}
    \Sigma & = \frac{1}{\tau} \int_{0}^{\tau}\|H_d\left(t\right)\| \mathrm{d} t,
\label{cost}
\end{align}
where $\|H_d\left(t\right)\|$ denotes the operator norm of $H_d\left(t\right)$. In our experiment, each unitary step is implemented over a fixed duration of $80\,\mathrm{ns}$. The integration in Eq.~\eqref{cost} can be written as $ \int_{0}^{\tau}g_dv\left(t\right) \mathrm{d}t$, which corresponds precisely to the definition of the Rabi angle $\theta$ and is independent of the pulse envelope. 

Combining thermodynamic cost with the energy change $\Delta E$, we define the thermodynamic efficiency as $\eta = \Delta  E / (\Delta  E + \Sigma)$. For the state preparation step, $\Delta E$ denotes the energy stored; for the extraction steps, it represents the amount of extracted work. Figure~\ref{fig4} shows the efficiencies of $R_Y(\theta)$, $V_\pi$, $U_c$, and total extraction processes (Total) as functions of $C(\rho_\theta)$. The corresponding Rabi angles range from $\pi/2$ to $\pi$. The efficiency of coherent ergotropy extraction increases monotonically with coherence, since coherent ergotropy grows with coherence while the thermodynamic cost decreases. Conversely, the efficiencies of incoherent and total ergotropy extraction processes decrease with increasing coherence.

Three features emerge from the results. First, the preparation efficiency exhibits a local maximum at a Rabi angle $\theta_m \approx 0.742\pi$ and coherence $C_m \approx 0.432$, marked by a red star in the inset. This indicates an optimal energy-to-cost ratio for charging the QB in the absence of decoherence. Combined with the results of Fig.~\ref{fig3}, this suggests a trade-off between maximum efficiency and maximum ergotropy lifetime in realistic battery operation. Second, at $\theta_e \approx 0.722\pi$, the efficiencies of coherent and incoherent ergotropy extraction are equal, corresponding to coherence $C_e \approx 0.
469$, marked by a yellow star in the inset. Notably, this does not imply equal amounts of $\mathcal{E}_c$ and $\mathcal{E}_i$, but rather equal efficiency due to the combined effects of work and thermodynamic cost. Third, when coherence reaches its maximum or minimum, the coherent (incoherent) ergotropy extraction efficiency matches that of state preparation, forming a closed curve in the efficiency-coherence plane. This occurs because in these limits, all ergotropy is coherent (incoherent), and the corresponding operations $U_C$ ($V_\pi$) are the inverse of $R_Y(\theta)$.

In summary, we have experimentally resolved coherent and incoherent ergotropy in a superconducting transmon qubit using three distinct work extraction protocols. Our results provide the first experimental verification on a superconducting qubit platform that, while quantum coherence does not alter the energy of a state, it fundamentally determines the amount of extractable work~\cite{korzekwa2016,francica2019a,cakmak2020,francica2020}. By varying the Rabi angle of the initial state, we observe a clear positive correlation between coherence and coherent ergotropy, and a negative correlation for incoherent ergotropy. This suggests that the choice of initial state can be optimized based on the decoherence characteristics of the QBs—for instance, storing coherent ergotropy to avoid energy relaxation or storing incoherent ergotropy to avoid dephasing. Furthermore, by analyzing the thermodynamic cost and efficiency of each step in the extraction cycle, we identify an optimal initial state that maximizes the charging efficiency. Altogether, our findings highlight a fundamental trade-off between coherence and efficiency in quantum energy storage, offering practical guidance for improving the performance of quantum devices. Future work could extend these concepts to multi-qubit systems, exploring how coherence-based strategies generalize in the presence of entanglement, laying the groundwork for scalable QBs. Additionally, we have systematically compared several methods of mixed-state preparation. These methods are platform-independent and, beyond quantum thermodynamics, can also be applied to quantum simulations of open systems~\cite{harrington2022} or non-Hermitian systems~\cite{lin2022}.

\section{ACKNOWLEDGMENTS}
This work was supported by the Micro/Nano Fabrication Laboratory of Synergetic Extreme Condition User Facility (SECUF). The devices were made at the Nanofabrication Facilities at the Institute of Physics, CAS in Beijing. This work was supported by Innovation Program for Quantum Science and Technology
(Grant No. 2021ZD0301800), the National Natural Science Foundation of China (Grants No. 12204528, No. 92265207, No. T2121001, No. 92065112, No. 92365301, No. T2322030, No. 12504593).
\bibliography{Experimental_Extraction_of_Coherent_Ergotropy}

\end{document}